\documentclass[prd,nofootinbib,superscriptaddress,showpacs]{revtex4}
\pdfoutput=1
\usepackage{graphicx}
\usepackage{epsfig}
\usepackage{latexsym}
\usepackage{amsmath}
\usepackage{amsfonts}
\usepackage{amsxtra}

\newcommand{\msbar}{\overline{\mbox{{\sc ms}}}}
\newcommand{\ams}{\alpha_{\msbar}}

\def\krto{ {\,\,\lower .8ex\hbox {$\longrightarrow \atop k \rightarrow 0$}\,\,}}

\def\bea{\begin{eqnarray} }
\def\beq{\begin{eqnarray} }

\def\eea{\end{eqnarray}}
\def\eeq{\end{eqnarray}}
\def\altura#1{\rule[0cm]{0cm}{#1cm}}
\def\eq#1{Eq.~(\ref{#1})}

\def\amsbar#1{\alpha_{\msbar}(#1)}

\newcommand{\lwrsim}{\raise0.3ex\hbox{$<$\kern-0.75em\raise-1.1ex\hbox{$\sim$}}}

\begin{document} 

\title{High statistics determination of the strong coupling constant in Taylor scheme and 
its OPE Wilson coefficient from lattice QCD with a dynamical charm}

\author{B.~Blossier}
\author{Ph.~Boucaud} 
\affiliation{Laboratoire de Physique Th\'eorique, 
Universit\'e de Paris XI; B\^atiment 210, 91405 Orsay Cedex; France}
\author{M.~Brinet}
\affiliation{Laboratoire de Physique Subatomique et de Cosmologie, CNRS/IN2P3/UJF; 
53, avenue des Martyrs, 38026 Grenoble, France}
\author{F.~De Soto}
\affiliation{Dpto. Sistemas F\'isicos, Qu\'imicos y Naturales, 
Univ. Pablo de Olavide, 41013 Sevilla, Spain}
\author{V.~Morenas}
\affiliation{Laboratoire de Physique Corpusculaire, Universit\'e Blaise Pascal, CNRS/IN2P3 
63177 Aubi\`ere Cedex, France}
\author{O.~P\`ene}
\affiliation{Laboratoire de Physique Th\'eorique, 
Universit\'e de Paris XI; B\^atiment 210, 91405 Orsay Cedex; France}
\author{K.~Petrov}
\affiliation{Laboratoire de
l'Acc\'el\'erateur Lin\'eaire, Centre Scientifique d'Orsay; B\^atiment 200, 91898 ORSAY Cedex, France}
\author{J.~Rodr\'{\i}guez-Quintero}
\affiliation{Dpto. F\'isica Aplicada, Fac. Ciencias Experimentales; 
Universidad de Huelva, 21071 Huelva; Spain.}
\affiliation{CAFPE, Universidad de Granada, E-18071 Granada, Spain}

\begin{abstract}

This paper reports on the determination of $\alpha_S$ from lattice simulations with 
2+1+1 twisted-mass dynamical flavours {\it via} the computation of the ghost-gluon coupling renormalized 
in the MOM Taylor scheme. A high-statistics sample of gauge configurations, used to evaluate the coupling from 
ghost and gluon propagators, allows for the appropriate update of previous results, now performing an improved 
analysis of data with reduced statistical errors and the systematical uncertainties under a better control.

\end{abstract}

\pacs{12.38.Aw, 12.38.Lg}

\maketitle

\begin{flushright}
LPT-Orsay 13-77\\
UHU-FT/13-09 \\
\end{flushright}
\vspace*{-1cm}
\begin{figure}[h]
    \includegraphics[width=25mm]{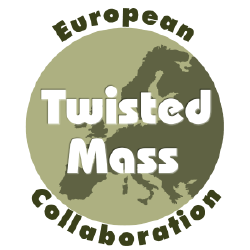}
\end{figure}




\section{Introduction}

With QCD in our hands as the theory for the strong interaction, the running of the  strong coupling 
constant can be computed by  the perturbation theory in the UV domain, supplemented by 
 nonperturbative OPE corrections when the IR domain is  approached. 
Thus,  values of the strong coupling constant obtained from experimental data at many different scales 
can be confronted to each others and to the theoretical running.
This successful confrontation is a benchmark calculation of QCD. 
When integrated, the Renormalization Group equation which governs the running of the coupling constant introduces
an energy scale,  $\Lambda_{\rm QCD}$, as a boundary condition.
$\Lambda_{\rm QCD}$ is defined for a given  number of active quark species and can be only obtained as an input from experiment.

Lattice computations can be used for the purpose of deriving $\Lambda_{\rm QCD}$ 
from an experimental input. The usual procedure can be shortly described as follows.  
Any quantity extracted from  a lattice simulation is obtained in unit of the lattice spacing, $a$.
This scale depends on the bare coupling and others bare set-up parameters  used in the simulation.
Its  value in physical unit is fixed by confronting  a low-energy experimental number (a mass, a decay
constant, $\ldots$) with its   lattice prediction.
Once the lattice spacing is known, the coupling at a given momentum 
can be computed from the lattice. Then, $\Lambda_{\rm QCD}$ 
can be subsequently derived from that coupling value. Among the methods most extensively applied 
to measure a coupling, one could find the implemention of the Schr\"odinger functional scheme (see,  for
instance,~\cite{Luscher:1993gh,deDivitiis:1994yp,DellaMorte:2004bc,Aoki:2009tf}
and  references therein), those based on the perturbative analysis of
short-distance sensitive lattice observables as the inter-quark static potential 
(see for instance~\cite{Booth:1992bm,Brambilla:2010pp}),  
the ``boosted'' lattice coupling (see~\cite{Gockeler:2005rv,Mason:2005zx,Maltman:2008bx,Davies:2008sw} and
references therein), moments of charmonia two-point correlation functions  (see~\cite{Allison:2008xk,McNeile:2010ji,Jansen:2011vr} and references therein)  or, in particular, those based on the study of  the
momentum behaviour of Green functions 
(see~\cite{Alles:1996ka,Boucaud:1998bq,Boucaud:2000ey,Boucaud:2000nd,Boucaud:2001st,Boucaud:2001qz,Sternbeck:2007br}, for instance). 

Very recently, in refs.~\cite{Blossier:2012ef,Blossier:2011tf}, we reported on our preliminary studies of  
the strong coupling running and evaluation of $\Lambda_{\rm QCD}$ with lattice simulations including 
up, down, strange and charm dynamical quarks in the sea. 
These two were the last of a series of works (see also \cite{Boucaud:2008gn,RodriguezQuintero:2009jw,Blossier:2010ky,Blossier:2010we}) aimed at computing the strong coupling through the lattice determination of the ghost-gluon 
coupling in the so-called MOM Taylor renormalization scheme. 
A similar program developed in parallel by different authors~\cite{Sternbeck:2007br,Sternbeck:2010xu,Sternbeck:2012qs}, 
although following a different approach for the data analysis, 
is also in progress. 
The current paper is entirely devoted to complete this series of papers with upgraded results  
obtained by performing an improved analysis. The main {\it ace} for the improvement is 
the use of a new high-statistics sample of gauge configurations which reduces 
the statistical errors and allows for a better control of systematic uncertainties. In particular, 
we relax  assumptions  previously made for the $O(4)$-invariant 
lattice artefacts which can be fitted from the data. Such an improved analysis confirms first  
that nonperturbative power corrections are required for the running description of the data
within our available momentum window. These power corrections, at the leading 
order\footnote{The OPE power corrections are found to be dominated by a non-vanishing 
landau-gauge dimension-two gluon condensate~\cite{Boucaud:2000nd,Gubarev:2000nz,Dudal:2005na,Boucaud:2002nc}}, 
appear to behave as predicted by SVZ sum-rules~\cite{Shifman:1978bx,Shifman:1978by}. 
Then, we finally provide with a lattice estimate for $\Lambda_{\rm QCD}$ in good agreement 
with the current ``{\it world average}" given by Particle Data Group~\cite{Beringer:1900zz}.


\section{$\boldsymbol{\alpha_S(q^2)}$ from the ghost-gluon coupling}
\label{sec:proc}

Updating the determination of the strong running coupling from the ghost-gluon vertex in the MOM Taylor scheme 
reported in \cite{Blossier:2012ef} is the main goal of this work. The details of the 
procedure, that will be hereupon outlined, have been profusely described in many previous works~\cite{Boucaud:2008gn,Blossier:2010ky,Blossier:2010we,Blossier:2011tf,Blossier:2013te}. 

The keystone for an accurate determination of $\alpha_S(q^2)$ from the ghost-gluon coupling in the MOM Taylor scheme 
is the well-known Taylor's statement about the non-renormalization of the ghost-gluon vertex with vanishing incoming 
ghost momentum in Landau gauge~\cite{Taylor:1971ff}. This implies that one can define a particular scheme for the 
QCD coupling such that its running with the renormalization momenta only relies on two-point Green functions renormalization 
constants~\cite{Sternbeck:2007br,Boucaud:2008gn,vonSmekal:1997vx},
\beq\label{alpha} 
\alpha_T(\mu^2) \equiv \frac{g^2_T(\mu^2)}{4 \pi}=  \ \lim_{\Lambda \to \infty} 
\frac{g_0^2(\Lambda^2)}{4 \pi} Z_3(\mu^2,\Lambda^2) \widetilde{Z}_3^{2}(\mu^2,\Lambda^2) \ ,
\eeq
where $Z_3$/$\widetilde{Z}_3$ stand for the gluon/ghost propagator renormalization constant and $\Lambda$ is a
non specified UV cut-off.  The main advantage resulting from this definition is that a very accurate nonperturbative estimate of the coupling and a very precise knowledge for its running from continuum perturbation tools can be both attained. 
It is worth to recall that an effective QCD charge definition, related to the pinching-technique effective 
charge~\cite{Aguilar:2008fh} and useful for phenomenological purposes~\cite{Binosi:2009qm}, can be grounded on this coupling in MOM Taylor scheme~\cite{Aguilar:2009nf,Aguilar:2010gm}. 
\eq{alpha} being regularization independent, we will use lattice-regularized ghost and gluon 
propagators to estimate $\alpha_T$, provided that the lattice artefacts are properly kept under control such that the 
continuum limit is well recovered. The latter will be tested by considering, as will be seen below, different lattice set-up's 
and verifying that \eq{alpha}'s results do not depend on the UV regularization cut-off, {\it i.e.} the lattice spacing.

\subsection{The running of $\boldsymbol{\alpha_T(q^2)}$}

Specially crucial is the very precise knowledge for the running of $\alpha_T$ with momenta. 
At the four-loop level, from continuum perturbation theory, and with an Operator Product expansion (OPE) at the leading power correction, it reads~\cite{Blossier:2010ky}, 
\beq\label{alphahNP}
\alpha_T(\mu^2)
\ = \
\alpha^{\rm pert}_T(\mu^2)
\ 
\left( 
 1 + \frac{9}{\mu^2} \
R\left(\alpha^{\rm pert}_T(\mu^2),\alpha^{\rm pert}_T(q_0^2) \right) 
\left( \frac{\alpha^{\rm pert}_T(\mu^2)}{\alpha^{\rm pert}_T(q_0^2)}
\right)^{1-\gamma_0^{A^2}/\beta_0} 
\frac{g^2_T(q_0^2) \langle A^2 \rangle_{R,q_0^2}} {4 (N_C^2-1)}
 + o\left(\frac 1 {\mu^2}\right) \right) \, \nonumber \\
\eeq
where $q_0$ is a renormalization point to be fixed, $\gamma_0^{A^2}$ can 
be taken from \cite{Gracey:2002yt,Chetyrkin:2009kh} and gives for $N_f=4$, 
\beq
1-\gamma_0^{A^2}/\beta_0 = \frac {27}{132 - 8 N_f} = \frac{27}{100} \ ;
\eeq
and, taking advantage of the $\msbar$ Wilson coefficients for the gluon and ghost OPE expansions 
at the ${\cal O}(\alpha^4)$-order~\cite{Chetyrkin:2009kh}, 
one can obtain in the appropriate renormalization scheme~\cite{Blossier:2013te}
\beq
R\left(\alpha,\alpha_0\right) = 
\left( 1 + 1.03735 \alpha + 1.07203 \alpha^2 + 1.59654 \alpha^3 \right) 
\ 
\left( 1 - 0.54993 \alpha_0 - 0.14352 \alpha_0^2 - 0.07339 \alpha_0^3 \right) \ ,
\eeq
to be plugged into \eq{alphahNP}. 
The purely perturbative running in \eq{alphahNP} is given up to four-loops by~\cite{Nakamura:2010zzi}
\begin{align}
  \label{betainvert}
  \begin{split}
      \alpha_T^{\rm pert}(\mu^2) &= \frac{4 \pi}{\beta_{0}t}
      \left(1 - \frac{\beta_{1}}{\beta_{0}^{2}}\frac{\log(t)}{t}
     + \frac{\beta_{1}^{2}}{\beta_{0}^{4}}
       \frac{1}{t^{2}}\left(\left(\log(t)-\frac{1}{2}\right)^{2}
     +
\frac{\widetilde{\beta}_{2}\beta_{0}}{\beta_{1}^{2}}-\frac{5}{4}\right)\right)
\\
     &+ \frac{1}{(\beta_{0}t)^{4}}
 \left(\frac{\widetilde{\beta}_{3}}{2\beta_{0}}+
   \frac{1}{2}\left(\frac{\beta_{1}}{\beta_{0}}\right)^{3}
   \left(-2\log^{3}(t)+5\log^{2}(t)+
\left(4-6\frac{\widetilde{\beta}_{2}\beta_{0}}{\beta_{1}^{2}}\right)\log(t)-1\right)\right)
   \end{split}
\end{align}
with $t=\ln \frac{\mu^2}{\Lambda_T^2}$ and the coefficients of the $\beta$-function
in Taylor-scheme~\cite{Chetyrkin:2000dq} (see also \cite{Boucaud:2008gn} where these coefficients have been 
shown to result from  those for the ghost and gluon anomalous dimensions). 
 The $\Lambda_{\rm QCD}$ parameters in Taylor-scheme and $\overline{\rm MS}$ are  related
through~\cite{Blossier:2010ky}
 \beq\label{ratTMS}
 \frac{\Lambda_{\overline{\rm MS}}}{\Lambda_T} \ = 
 e^{\displaystyle - \frac{507-40 N_f}{792 - 48 N_f}} \ = \ 0.560832 
 \ ,
 \eeq
for the ${\rm N}_f=4$ case. Thus, an accurate determination of $\Lambda_{\msbar}$ and the $\msbar$ 
version of \eq{betainvert} allows for obtaining the standard $\msbar$ running coupling at any momentum. 

The accurate determination of $\Lambda_{\msbar}$ will result, as will be seen below, from the confrontation of 
\eq{alphahNP} with a lattice estimate of \eq{alpha} within the appropriate momenta window. 
The ghost and gluon renormalization constants to be here applied in \eq{alpha} have been   
obtained from $N_f=2+1+1$ gauge configurations for several bare couplings, light twisted masses and volumes. 
This allows for a thorough study of the artefacts from the lattice regularization and the required 
chiral extrapolation. The first type of artefacts affecting the measured quantity $\alpha_T^{\rm Latt}$ concerns 
the breaking of the full $O(4)$ rotational symmetry. The the so-called $H(4)$-extrapolation  procedure~\cite{Becirevic:1999uc,Becirevic:1999hj,deSoto:2007ht}
exploits the remaining $H(4)$ isometry group symmetry  and the discrepancies within $H(4)$ orbits,
which at small enough momenta can be written as :
\beq\label{eq:H4}
\alpha_T^{\rm Latt}\left(a^2q^2,a^2 \frac{q^{[4]}}{q^2},\dots \right) 
\ = \ \widehat{\alpha}_T(a^2 q^2) \ + \ 
\left. \frac{\partial \alpha_T^{\rm Latt}}{\partial \left(a^2
\frac{q^{[4]}}{q^2}\right)}  
\right|_{a^2 \frac{q^{[4]}}{q^2}=0} \ 
a^2 \frac{q^{[4]}}{q^2} \ + \ \dots
\eeq
where $q^{[4]}=\sum_i q_i^4$ is the first relevant  $H(4)$-invariant, to cure  these effects  (see appendix \ref{sec:apH4}).
Then $\widehat{\alpha}_T$ is written as :
\beq\label{eq:a2p2}
\widehat{\alpha}_T(a^2 q^2) \ = \ \alpha_T(q^2) \ + \ c_{a2p2} \ a^2 q^2 \ + \ o(a^2) \ ,
\eeq
 which can be used  to fit and eliminate the dominant  $O(4)$-invariant artefact 
from the lattice estimates~\cite{Boucaud:2003dx,Boucaud:2005rm,Blossier:2010vt}.  
One is then left with the continuum prediction, $\alpha_T$, to be confronted with \eq{alphahNP}.


\subsection{The lattice data}
\label{subsec:lat}

\begin{figure}[h]
  \begin{center}
    \includegraphics[width=12cm]{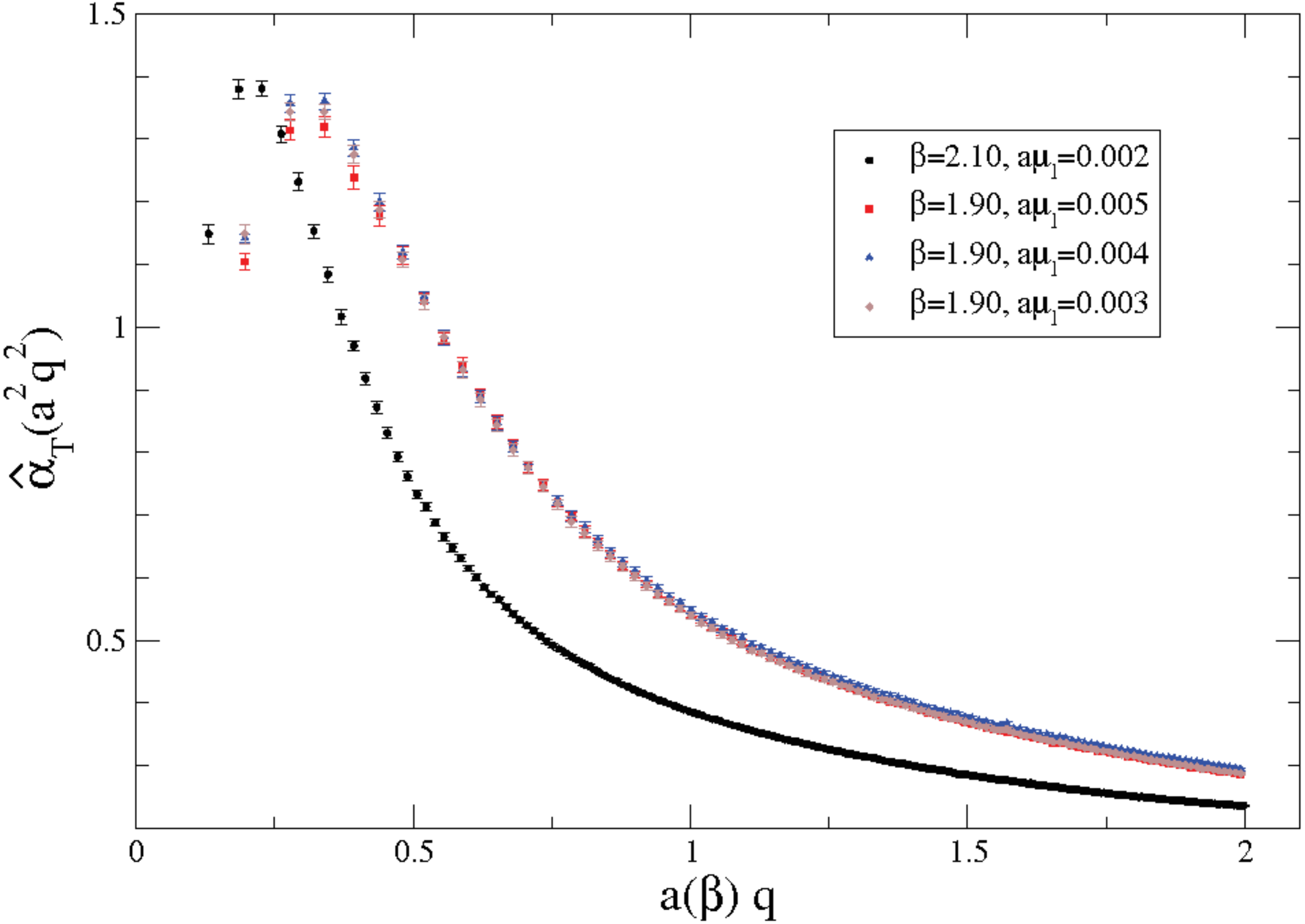}
  \end{center}
\caption{\small Lattice estimates for $\hat{\alpha}_T$ from the gluon and ghost propagators 
obtained with the ensembles of Tab.~\ref{tab:set-up}, applied first in \eq{alpha} and cured then 
for the $H(4)$-invariant hypercubic artefacts as explained around \eq{eq:H4}. They appear 
plotted in terms of the lattice momenta, $a(\beta) q$, and a first cut, $a(\beta) q < 2$, is 
applied to avoid large hypercubic artefacts.
}
\label{fig:raw}
\end{figure}

Then, Eqs.~(\ref{alphahNP}-\ref{eq:a2p2}) can be used to predict the running of $\alpha_T(q^2)$ 
for a momenta window where OPE higher-powers, $o(1/q^2)$, and 
$o(a^2q^2)$ lattice artefacts could be neglected. Within such a window, the only three parameters, 
$g^2 \langle A^2 \rangle$,  $\Lambda_{\overline{\rm MS}}$ and the coefficient for the $O(4)$-invariant
artefacts $c_{a2p2}$, remain free to be fitted to account for the lattice estimate of the Taylor coupling through 
\eq{alpha}. Aiming at obtaining $\alpha_T^{\rm Latt}$ by 
\eq{alpha}, the ghost and gluon propagators are computed from the gauge configurations 
simulated at several lattices with $N_f=2+1+1$ twisted-mass lattice flavors~\cite{Frezzotti:2000nk} by 
the ETM collaboration~\cite{Baron:2010bv,Baron:2011sf}. In the gauge sector, we use 
the Iwasaki action and compute the propagators as described in
refs.~\cite{Blossier:2010ky}, 
while for the fermion action we have
\beq\label{eq:tmSl}
S_l \ = \ a^4 \sum_x \overline{\chi}_l(x) \left( \altura{0.5} D_W[U] + m_{0,l} + i
\mu_l \gamma_5 \tau_3 \right) \chi_l(x)
\eeq
for the doublet of degenerate light quarks~\cite{Frezzotti:2003xj} and
\beq\label{eq:tmSh}
S_h \ = \ a^4 \sum_x \overline{\chi}_h(x) \left( \altura{0.5} D_W[U] + m_{0,h} 
+ i \mu_\sigma \gamma_5 \tau_1 + \mu_\delta \tau_3 \right) \chi_h(x)
\eeq
for the heavy doublet. $D_W[U]$ is the standard massless Wilson Dirac operator. 
The lattice parameters for the  ensembles of gauge
configurations we 
used are given in tab.~\ref{tab:set-up}. 
Tuning to maximal twist is achieved by choosing a parity odd operator
and determining $\kappa_{crit}$ such that this operator has a vanishing expectation 
value. One appropriate quantity is the PCAC light quark mass and we demand
$m_{PCAC}$ = 0.
We refer the interested reader to
refs.~\cite{Baron:2010bv,Baron:2011sf} 
for more details about the set-up of the twisted mass lattice simulations. 
The results for $\widehat{\alpha}_T$ obtained, as explained, with \eq{alpha} from 
the lattice data for the ensembles in Tab.~\ref{tab:set-up} appear plotted in 
Fig.~\ref{fig:raw}. The errors have been estimated through the usual jackknife method, 
by applying the $H(4)$-extrapolation procedure, explained above, to the estimates of $\widehat{\alpha}_T$ 
from each jackknife cluster of configurations we made.
As can be seen,  this $H(4)$-extrapolation works pretty well and  
very smooth curves, with no trace of the usual oscillations from the $O(4)$ breaking, 
are obtained. 

\begin{table}[ht]
\begin{tabular}{||c|c|c|c|c|c|c||}
\hline
$\beta$ & $\kappa_{\rm crit}$ & $a \mu_l$ & $a \mu_\sigma$ & $a \mu_\delta$ &
$(L/a)^3\times T/a$ & confs. \\
\hline
1.90 & 0.163267 & 0.0050 & 0.150 & 0.190  & $32^3\times 64$ & 500 \\
     & 0.163270 & 0.0040 &           &           & $32^3\times 64$ & 500 \\
     & 0.163272 & 0.0030 &           &            & $32^3\times 64$ & 500 \\   
\hline
2.1 &  0.156357 & 0.0020 & 0.120 & 0.1385 & $48^3\times 96$ & 800 \\  
\hline
\end{tabular}
\caption{Lattice set-up parameters for the ensembles we used in this paper. 
The last column stands for the number of gauge field configurations we used.}
\label{tab:set-up}
\end{table}

\section{Analysing the data}

The first step in our strategy for analysing the data is to exploit the high statistics (800 gauge configurations) 
for the lattice ensemble at $\beta=2.10$ (see Tab.~\ref{tab:set-up}) in order to extract precisely the physical 
running of $\alpha_T(q^2)$.  Eqs.~(\ref{alphahNP}-\ref{eq:a2p2}), obtained from continuum perturbation theory 
and OPE analysis, are thought to describe the running behaviour of the lattice estimate for 
the Taylor coupling through \eq{alpha}, only up to $o(1/q^2)$ higher-power OPE corrections 
and $o(a^2q^2)$ lattice artefacts. Then, in order to apply Eqs.~(\ref{alphahNP}-\ref{eq:a2p2}) and determine 
its three free-parameters by a fit, one needs to identify the appropriate window of momenta, large enough for 
the running behaviour not to be polluted by higher OPE powers but small enough as not to be affected 
by higher-order discretization artefacts. As far as we are dealing with only one lattice ensemble, at 
given bare coupling ($\beta$) and quark masses ($a\mu_l$, $a\mu_\sigma$, $a\mu_\delta$), we can skip 
the problem of the absolute lattice calibration (the determination of the lattice spacing, $a(\beta)$, 
in physical units) that can be put back to a final stage of the analysis. Consequently, all the fitted dimensionful 
parameters will be now expressed in units of the lattice spacing and, in particular, the perturbative running given 
by \eq{betainvert} will depend on $t=\ln{(a^2(\beta)q^2/a^2(\beta)\Lambda^2_T)}$.

\subsection{The optimal window fit}

We first perform a fit, for every set of data with lattice momenta above $a(\beta) q \simeq 1.15$, whose results 
read as in Tab.~\ref{tab:opt}. Then, the upper and lower bounds for the fitting window have been systematically 
shifted, the window size roughly ranging from 0.4 to 0.8 in units of $a(2.10)^{-1}$, to look for the optimal fit 
(the one with the minimum value of $\chi^2$/d.o.f.). This optimal fit is found to happen 
for data with lattice momenta within $1.31 < a(\beta) q < 1.81$ (see Tab.~\ref{tab:opt}). In both cases, the renormalization 
point for \eq{alphahNP} is chosen\footnote{This choice would imply, had we taken the lattice spacing for $\beta=2.10$ 
used in ref.~\cite{Blossier:2011tf}, $q_0=10$ GeV.} to be $a(2.10)q_0=2.92$. 
 
\begin{table}[ht]
\begin{tabular}{||c|c|c|c|c||}
\hline
fit window & $\chi^2/{\rm d.o.f.}$ & $\Lambda_{\msbar} a(2.1)$ & $g^2 \langle A^2 \rangle a^2(2.1)$ & $c_{\rm a2p2}$ \\
\hline
[1.15,2.0] & 0.864 & 0.092(5) & 0.39(11) & -0.0049(8) \\
\hline
[1.31,1.81] & 0.270 & 0.099(3) & 0.26(7) & -0.0066(4)\\
\hline
\end{tabular}
\caption{Fitting parameters expressed in units of $a^{-1}(2.1)$ for the two different fitting windows discussed in 
the text: all momenta larger than $a(2.1)q > 1.15$ (first line) and the window for the optimal fit (second line). The errors 
reported here are statistical ones, obtained from the jackknife's method.}
\label{tab:opt}
\end{table}

\begin{figure}[ht]
  \begin{center}
    \includegraphics[width=12cm]{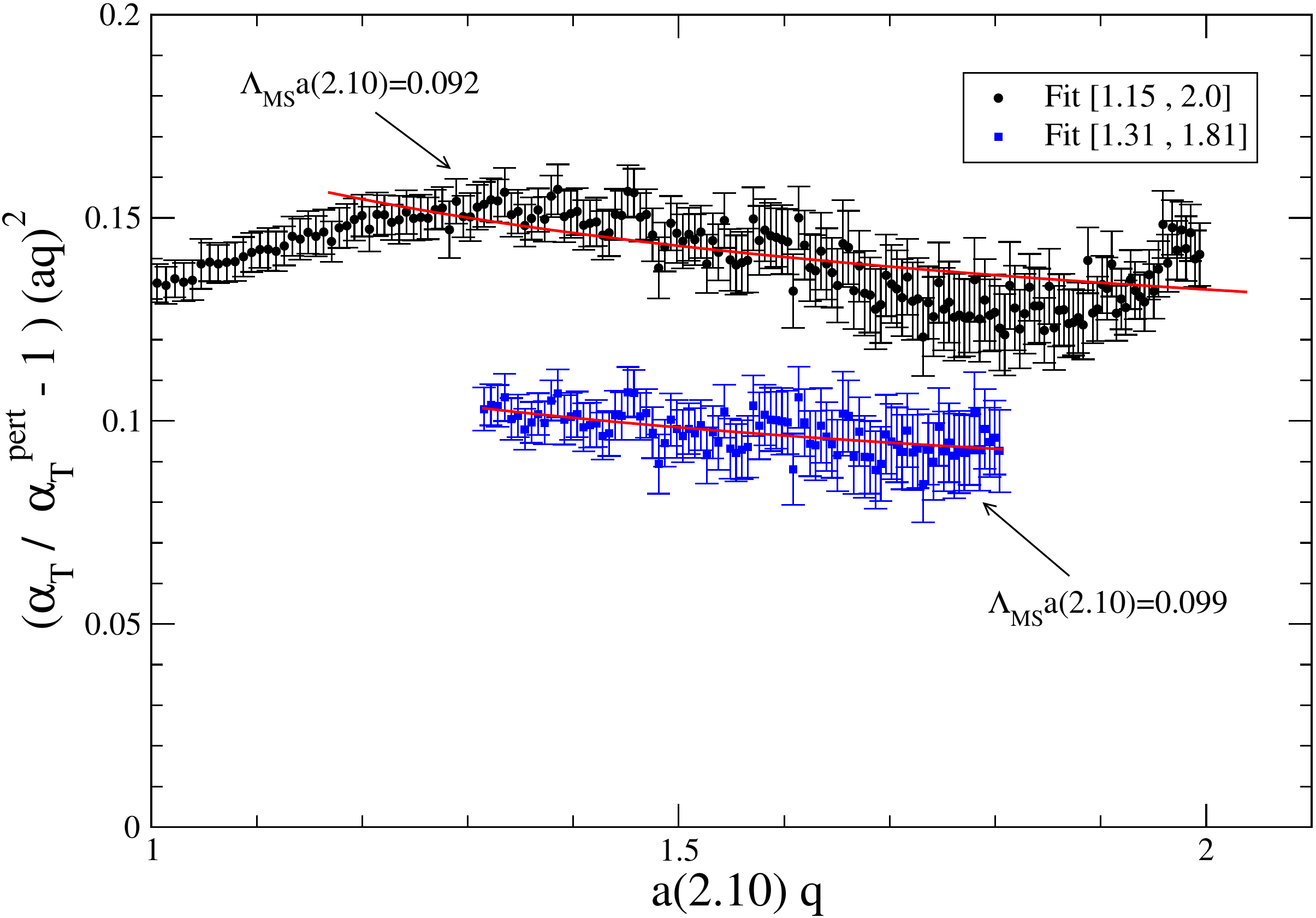}
  \end{center}
\caption{\small \eq{eq:mona}'s l.h.s. estimated  with $\alpha_T$ from the lattice data for the ensemble at $\beta=2.10$ (800 confs.) and its four-loop perturbative prediction evaluated with $\Lambda_{\msbar}$ from Tab.~\ref{tab:opt}. The solid red line corresponds 
to the evaluation of \eq{eq:mona}'s r.h.s., also with the best-fit parameter for $g^2\langle A^2 \rangle$ in  Tab.~\ref{tab:opt}. 
All the dimensionful parameters should be taken in units of the appropriate lattice spacing power.
}
\label{fig:NP}
\end{figure}

Furthermore, according to \eq{alphahNP}, 
\beq\label{eq:mona}
\left( \frac{\alpha_T(q^2)}{\alpha_T^{\rm pert}(q^2)} - 1 \right) \ q^2 \ = \  
9 \
R\left(\alpha^{\rm pert}_T(q^2),\alpha^{\rm pert}_T(q_0^2) \right) 
\left( \frac{\alpha^{\rm pert}_T(q^2)}{\alpha^{\rm pert}_T(q_0^2)}
\right)^{1-\gamma_0^{A^2}/\beta_0} 
\frac{g^2_T(q_0^2) \langle A^2 \rangle_{R,q_0^2}} {4 (N_C^2-1)}\ .
\eeq
up to terms vanishing at large $q^2$. Then, \eq{eq:mona}'s l.h.s. can be computed with $\alpha_T$ from the lattice through \eq{eq:a2p2} 
and its perturbative four-loop prediction from \eq{betainvert}, with the best-fit parameters 
for $\Lambda_{\msbar}$ from Tab.~\ref{tab:opt}.  As \eq{eq:mona}'s r.h.s. reads, 
this, plotted in terms of momenta in lattice units within the appropriate range, makes essentially 
the running of the Wilson coefficient to appear.  
This can be seen in Fig.~\ref{fig:NP}: a clearly nonzero nonperturbative 
contribution appears remarkably to behave as the OPE analysis predicts\footnote{A similar 
analysis has been also preliminary performed in \cite{Blossier:2011tf} where, with a poorer statistics 
for the lattice ensembles (50 gauge configurations), the Wilson coefficient running has also been  
proven to be required in order to properly describe the lattice data of $\alpha_T$ for physical 
momenta above around 4 GeV.}, although some systematic deviation for the data can be noticed from 
the expected behaviour in the case of the fit over every momenta above $a(2.1) q > 1.15$. 
The drastic diminution of $\chi^2/$d.o.f. when the fit is restricted to the optimal window gives also a 
strong indication that, within that window, higher-order OPE corrections and discretization 
artefacts can be properly neglected.

\subsection{Higher-order power corrections}

\begin{figure}[ht]
  \begin{center}
    \includegraphics[width=12cm]{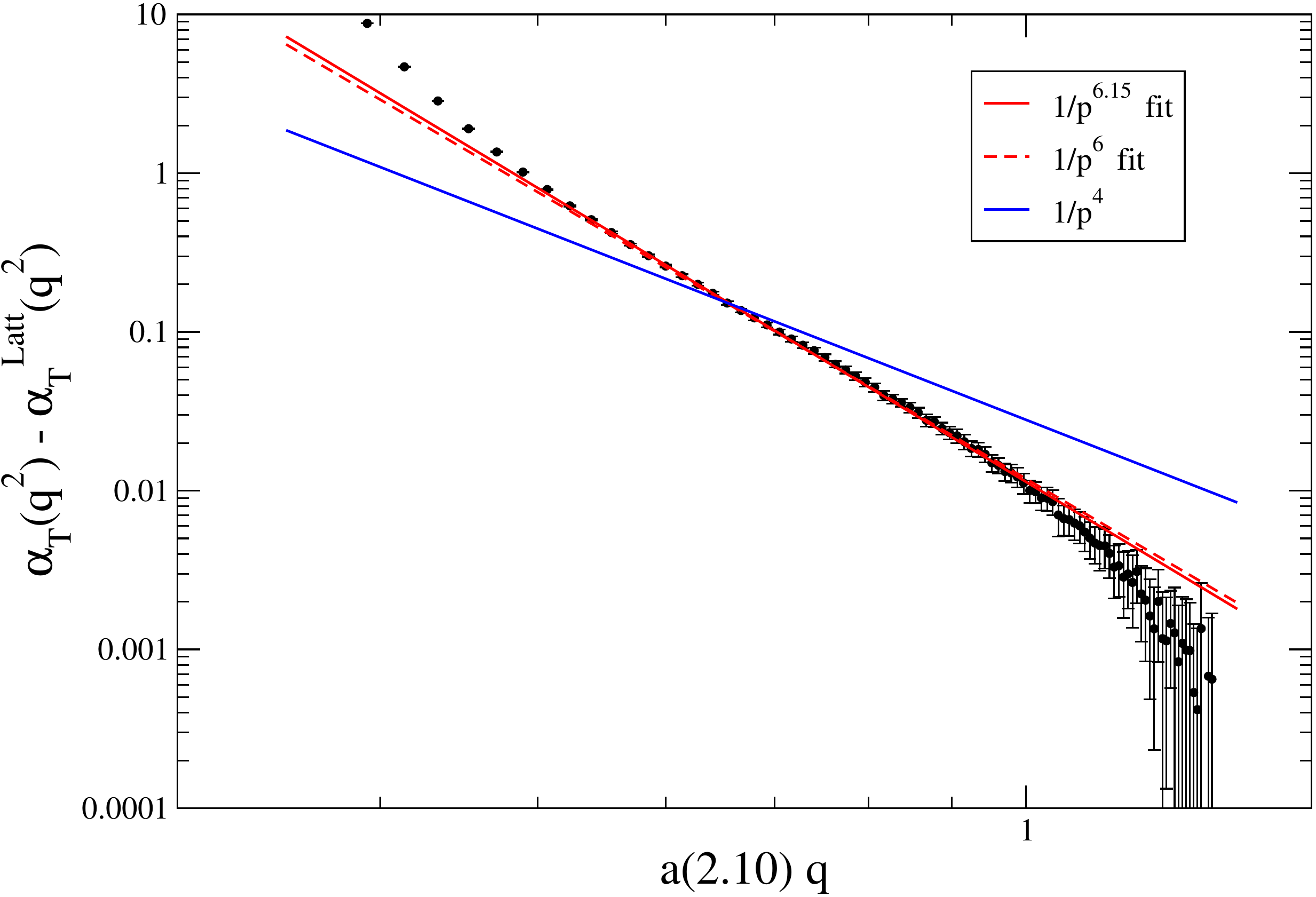}
  \end{center}
\caption{\small Lattice data deviations from \eq{alphahNP} with the best-fit parameters for the small window 
in Tab.~\ref{tab:opt}, plotted in terms of the momentum in units of the lattice spacing. Logarithmic scales 
are applied for the coordinates in both axes. A best-fit for the logarithmic slope of 6.15 is shown in red solid 
line as well as, for the sake of comparison, fits with slopes fixed to be 6 (red dashed) and 4 (blue).
}
\label{fig:HOPE}
\end{figure}

The Taylor coupling estimated from the lattice should feel at low momenta the impact of nonpertubative contributions 
other than the leading OPE ones included in \eq{alphahNP}. Indeed, lattice data deviations from this OPE formula are 
clearly visible for momenta $a(2.10)q~\lwrsim~1.2$ in fig.~\ref{fig:NP}. With the OPE machinery at hand, corrections 
to \eq{alphahNP} should be incorporated through condensates of higher-order local operators and appear thus suppressed 
by powers of the inverse of momentum higher than 2. Aiming at identifying the dominant next-to-leading contribution, 
lattice data deviations from \eq{alphahNP} are plotted in fig.~\ref{fig:HOPE} in terms of the momenta, using for both 
axes logarithmic scales. As can be seen, a logarithmic slope of $\sim 6.15$ results from lattice data, after the subtraction of 
\eq{alphahNP} with the best-fit parameters for the small window in Tab.~\ref{tab:opt}, for momenta above 
$a(2.10)q\sim 0.5$ and below $a(2.10)q\sim 1.3$. 

This logarithmic slope is consistent with the results we obtained in refs.~\cite{Blossier:2011tf,Blossier:2012ef}, where a $1/p^6$-correction is incorporated to describe successfully the 
lattice data all over a large momenta window roughly ranging from 1.75 GeV up to 7 GeV. As was discussed therein, 
such a $1/p^6$-correction might not be necessarily explained by the dominance of a condensate of dimension six but 
might be an effective power originated either by the interplay of a next-to-leading power correction and its Wilson coefficient 
or by a different nonperturbative mechanism being dominant at these low-momenta. Anyhow, on the ground of the striking 
result shown by Fig.~\ref{fig:HOPE}, one can firmly argue that \eq{alphahNP} with the addition of a $1/p^x$-contribution, where 
$x$ is around 6, should account for the running of lattice data also at very low momenta. Thus, \eq{eq:a2p2} could be replaced by 
\beq\label{eq:1overpx}
\widehat{\alpha}_T(a^2 q^2) \ =\ \alpha^{\rm A2}_T(q^2) \ + \ \frac{d_x}{q^x} \ + \ c_{a2p2} \ a^2 q^2 \ + o(a^2) \ ,
\eeq
where the physical running of the coupling is now given by 
\beq\label{eq:atA2}
\alpha_T(q^2) \ = \ \alpha_T^{A2}(q^2) \ + \ \frac{d_x}{q^x} \ ;
\eeq
 $\alpha^{A2}_T$ being still given by \eq{alphahNP}. 

We can now try to fit \eq{eq:1overpx} to the lattice estimates for the coupling through \eq{alpha}, for all momenta such that 
$a(2.10)q > 0.5$. However, if one keeps $c_{a2p2}$, $\Lambda_{\msbar}$, $g^2\langle A^2\rangle$, $d_x$ and the power $x$ 
as free parameters, all of them appear strongly correlated allowing for inconsistent values of $x$, with unnaturally large gluon condensates and much lower values of $\Lambda_{\msbar}$. Thus, as one can judiciously guess that higher-order power 
corrections may appear naturally correlated with the leading OPE contribution but should not affect too much  
the parameter driving the purely perturbative running, $\Lambda_{\msbar}$, we will apply the following two-step procedure: 
(i) first, we will take for $\Lambda_{\msbar}$ and $c_{a2p2}$ the values from Tab.~\ref{tab:opt} and fit  
\eq{eq:1overpx} to the data with $g^2\langle A^2\rangle$, $d_x$ and $x$ as free parameters; (ii) then, we will assume $x$, for the higher-order power corrections, to be well determined by the previous step and perform a new fit of \eq{eq:1overpx} to the data with the other four free parameters. The results are gathered in Tab.~\ref{tab:fitpw}. It should be noticed that 
a power value compatible with 6 within errors is found for the higher-order power correction  
and that the best-fit parameters here obtained and those in the previous section (without including higher-order 
corrections, but shifting the window lower bound up to a larger momentum) are also compatible within the errors.    
\begin{table}[h]
\begin{tabular}{|l|ccccc|}
\hline
&
$c_{a2p2}$ & 
$\Lambda_{\msbar} a(2.10)$ &
$g^2\langle A^2 \rangle a^2(2.10)$ & 
$d_x a^x(2.10)$ &
$x$ \\
\hline
step (i) & 
& &
0.29(2) & 
-0.0159(22) & 
{\bf 5.73(27)}
\\
\hline
step (ii)  &
{\bf -0.0058(4)} &
{\bf 0.0957(22)} &
{\bf 0.35(5)}& 
{\bf -0.0168(11)}&  
\\ \hline
\end{tabular}
\caption{\small Results obtained with the two-steps procedure, explained in the text, followed to account with \eq{eq:1overpx} for the Taylor coupling computed with the lattice at $\beta=2.10$. The bold-face characters indicate the best-fit parameters that will be applied in the following to describe data for any momenta above 
$a(2.10)q > 0.5$.}
\label{tab:fitpw}
\end{table}

\subsection{Higher-order lattice artefacts and the running of the Wilson coefficient}

\begin{figure}[ht]
  \begin{center}
    \includegraphics[width=12cm]{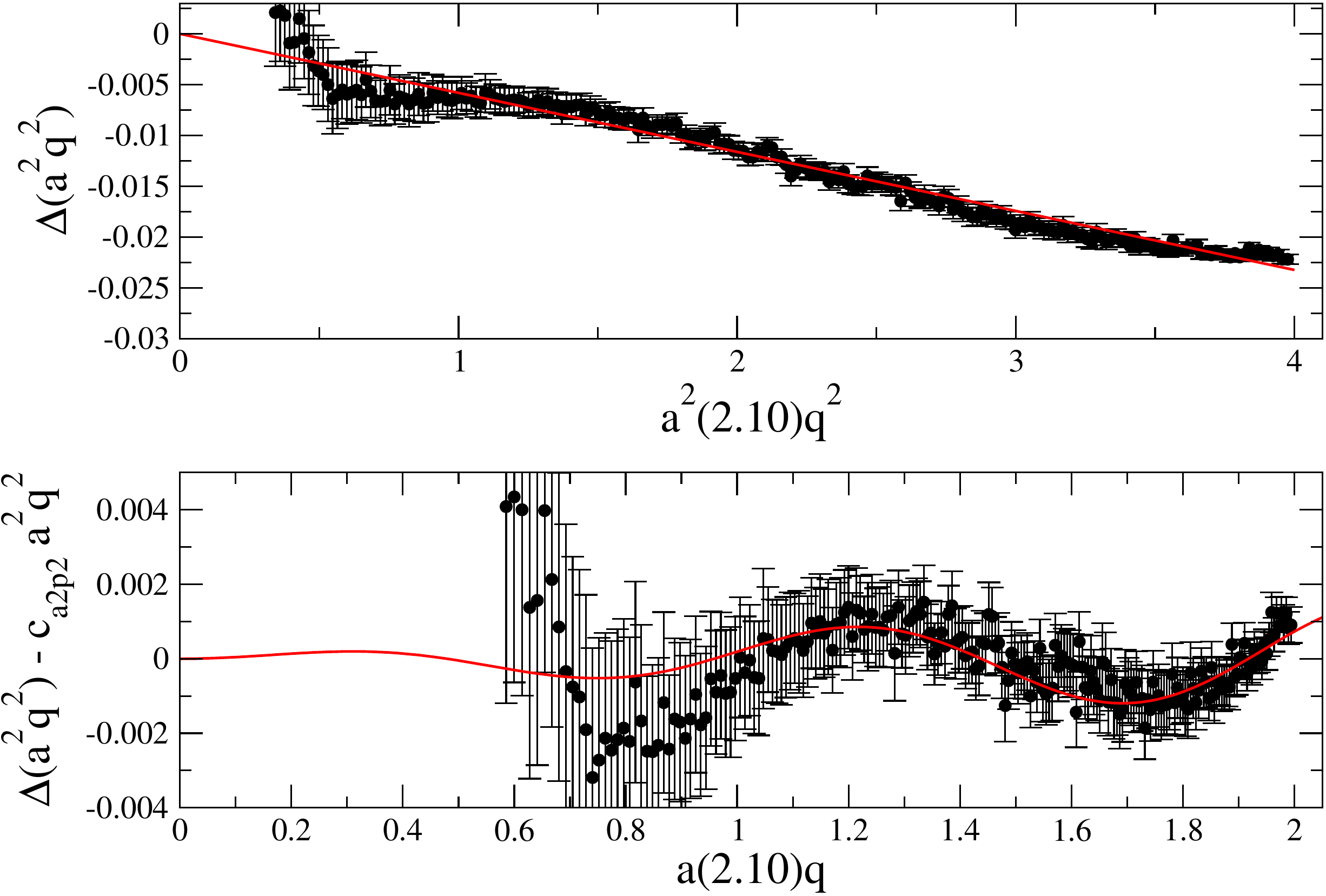}
  \end{center}
\caption{\small (Top) $\Delta(a^2q^2)$ computed through \eq{eq:Rvsa2q2} from lattice data after the subtraction of the running given 
by \eq{alphahNP} and the higher-order power correction. It is shown to behave as one expects 
due to $O(4)$-invariant leading artefacts (solid red line). Very small oscillatory deviations from the leading behaviour can be appreciated and appear plotted in terms of $aq$ (bottom) which can be fitted in practice by $r_0 aq \sin{(r_1 aq)}$ with 
$r_0=0.00071$ and $r_1=2.081 \, \pi$ (solid red line).
}
\label{fig:Rvsa2q2}
\end{figure}

According to \eq{eq:1overpx}, one can define 
\beq\label{eq:Rvsa2q2}
\Delta(a^2q^2) \ = \ \widehat{\alpha}_T(a^2q^2) - \alpha^{A2}_T(q^2) - \frac {d_x}{q^x} \ = \ c_{a2p2} \ a^2q^2 \ + \ o(a^2) \ ,
\eeq
that, up to $o(a^2)$-corrections, should behave linearly on $a^2q^2$. Indeed, we have found this to happen for 
$\Delta(a^2q^2)$ computed by subtracting $\alpha^{A2}_T(q^2)$ from \eq{alphahNP} and the higher-order power correction, with the best-fit parameters (in bold-faced characters) of Tab.~\ref{tab:fitpw}, to the coupling lattice data. The latter can be seen in  Fig.~\ref{fig:Rvsa2q2}. Furthermore, very small oscillatory deviations from the linear behaviour can be appreciated 
(bottom plot of Fig.~\ref{fig:Rvsa2q2}). This might imply for the $O(4)$-invariant lattice artefacts a more complicate structure, always behaving as $a^2q^2$ in a first order Taylor expansion, or a relic of $O(4)$-breaking artefacts not totally cured with the $H(4)$-extrapolation 
procedure\footnote{To remove $O(4)$-breaking artefacts at the order $O(a^2)$, one just needs to extrapolate $a^2q^{[4]}/q^2$ 
to 0. To go beyond, apart from $a^4(q^{[4]}/q^2)^2$, one would need to include $a^4q^{[6]}/q^2$ terms in the extrapolation which would so increase the difficulty of the computation.}. In practice, such a small deviation could be in practice pretty well parameterized by $r_0 aq \sin{(r_1 aq)}$, as can be seen in Fig.~\ref{fig:Rvsa2q2}.

\begin{figure}[ht]
  \begin{center}
    \includegraphics[width=12cm]{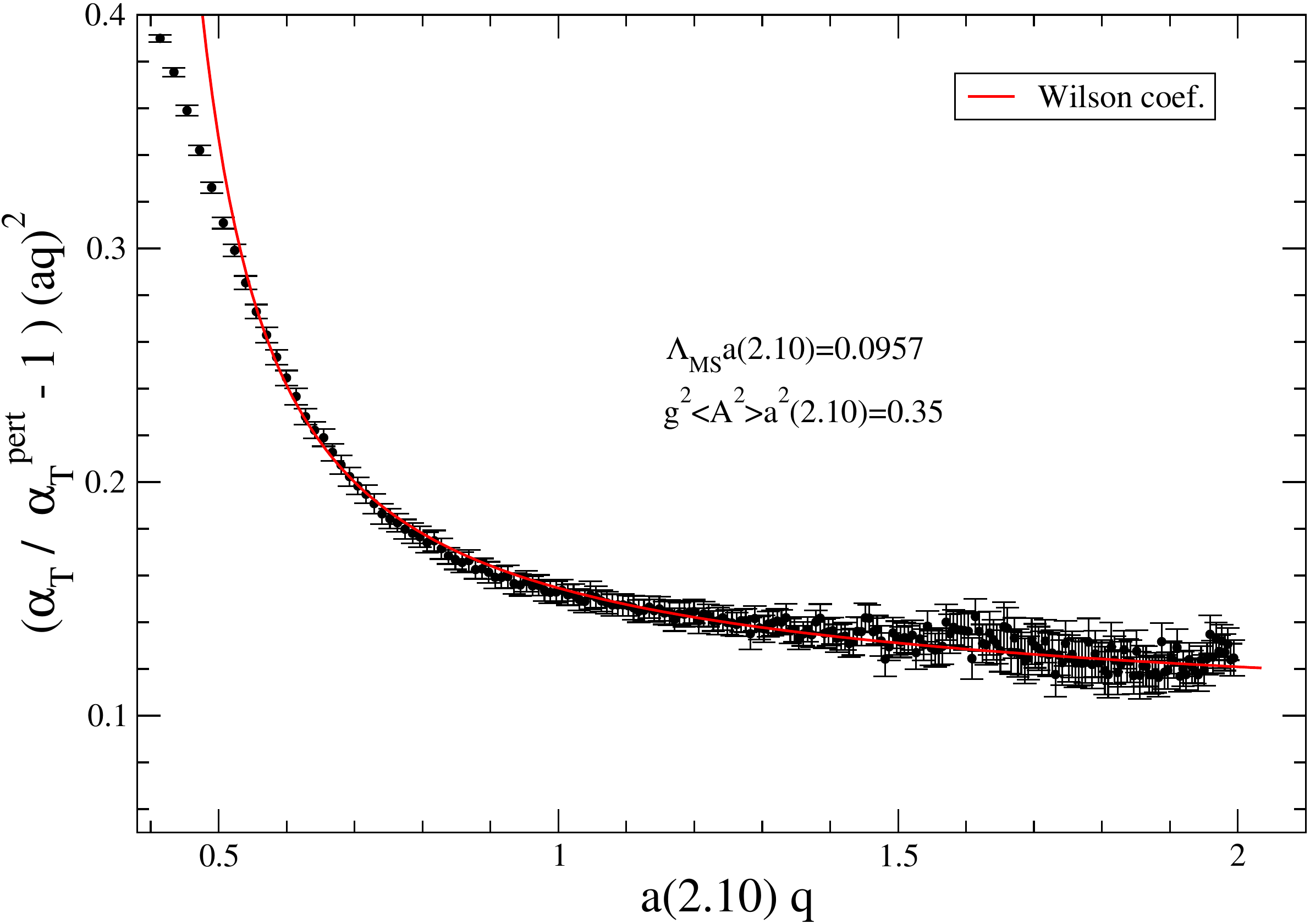}
  \end{center}
\caption{\small The same as plotted in Fig.~\ref{fig:NP} but $\alpha_T$ in \eq{eq:mona}'s l.h.s. is here evaluated only after 
dropping completely the lattice artefacts and the higher-order power correction away from the lattice data. The solid red line 
corresponds to the running predicted for the Wilson coefficient of the leading OPE power correction. 
}
\label{fig:NPtot}
\end{figure}

A final remark is now in order: after the removal of lattice artefacts and the higher-order power correction, the running 
of the leading OPE Wilson coefficient can be now properly shown over a large window, $a(2.10)q > 0.5$. 
In other words, \eq{eq:mona}'s l.h.s. can be now plotted in terms of $a q$ with $\alpha_T$ replaced by $\alpha^{A2}_T$ obtained from the lattice data through \eq{eq:1overpx} and compared with the theoretical expression given in \eq{eq:mona}'s  rhs. This is done in the plot of Fig.~\ref{fig:NPtot}, where the predicted running for 
the leading OPE Wilson coefficient is beautifully followed by the corrected lattice estimates for momenta above 
$a(2.10)q \sim 0.5$.

\section{Scaling and lattice calibration}

We have focussed until now the analysis in exploiting the high-statistics ensemble of data obtained for a $48^3 \times 96$ lattice 
at $\beta=2.10$ (see Tab.~\ref{tab:set-up}). We thus obtained the results collected in Tab.~\ref{tab:opt}, when fitting with Eqs.~(\ref{alphahNP},\ref{eq:a2p2}), and in Tab.~\ref{tab:fitpw}, when applying Eqs.~(\ref{alphahNP},\ref{eq:1overpx})  
including also a higher-order power correction. From now on, we can take advantage of the ``scaling" for the running of the 
physical coupling at different $\beta$'s in order to extract useful information, and perform a demanding crosscheck, from the three other 
ensembles of data at $\beta=1.90$ and different light masses (see Tab.~\ref{tab:set-up}).

\subsection{Scaling at different $\boldsymbol{\beta}$'s}

The running coupling is a renormalized quantity not depending on either the regularization scheme or the regularization cut-off. 
After being properly cured for lattice artefacts, the resulting coupling expressed in terms of the physical momentum should not depend on $\beta$ and, one can judiciously argue, only slightly on the mass of the active quark flavours, provided that the momentum runs far enough away from the quark mass thresholds\footnote{In the standard $\msbar$ approach, the flavours become active at momenta above their $\msbar$ running quark masses, with the appropriate radiative corrections included in the matching formula~\cite{Chetyrkin:2005ia,Schroder:2005hy,Beringer:1900zz}}. We will assume any flavour mass dependence 
to have been properly captured by the lattice spacing and thus write
\beq\label{eq:scaling}
\widehat{\alpha}^{(\beta,\mu_l)}_T(a^2(\beta,\mu_l)q^2) \ - \ c_{a2p2} \ a^2(\beta,\mu_l)q^2 
\ \equiv \
\widehat{\alpha}^{(\beta_0,\mu^0_l)}_T(a^2(\beta_0,\mu^0_l)q^2) \ - \ c_{a2p2} \ a^2(\beta_0,\mu^0_l)q^2 \ ,
\eeq
for any two bare coupling parameters, $\beta$ and $\beta_0$, and light flavour masses, $\mu_l$ and $\mu^0_l$ (we will neglect 
the effect of the heavy flavour masses). \eq{eq:scaling} stands for the explicit expression of the scaling condition for estimates from two different simulations at the same momentum $q$ in physical units. Now, we take $\beta_0=2.10$ and $\mu^0_l=0.002$ and, 
according to \eq{eq:1overpx}, replace \eq{eq:scaling}'s r.h.s. by the continuum running of $\alpha_T$ given by \eq{alphahNP} and 
the higher power correction, 
\beq\label{eq:scaling2}
\widehat{\alpha}^{(\beta,\mu_l)}_T\left( \left(\frac{a(\beta,\mu_l)}{a(\beta_0,\mu^0_l)}\right)^2 k^2_L \right) \ - \ c_{a2p2} \ \left(\frac{a(\beta,\mu_l)}{a(\beta_0,\mu^0_l)}\right)^2 k^2_L
&= &  \alpha_T(k^2_L;\Lambda_{\msbar}a(\beta_0,\mu^0_l),g^2\langle A^2\rangle a^2(\beta_0,\mu^0_l)) 
\nonumber \\
&+& \frac{d_x}{(a(\beta_0,\mu^0_l) k_L)^x} \ ;
\eeq
where $k_L=a(\beta_0,\mu^0_l)q$ stands for the momentum in units of the lattice spacing at $\beta_0$ and $\mu^0_l$. 
In \eq{eq:scaling2}'s r.h.s., we write down explicitly the parameters $\Lambda_{\msbar}$ and $g^2\langle A^2 \rangle$ 
used for $\alpha_T$ to specify that they are taken from Tab.~\ref{tab:fitpw}, also in units of the lattice spacing. 
It should be recalled that Eqs.~(\ref{alphahNP},\ref{eq:1overpx}) are only suitable for describing the lattice data at 
$\beta=2.10$ when $k_L > 0.5$. Then, this will be also the case for \eq{eq:scaling2}, although threshold mass effects 
might not be still negligible at $k_L \sim 0.5$.

\eq{eq:scaling2}'s r.h.s. is completely determined by \eq{alphahNP} and the best-fit parameters of Tab.~\ref{tab:fitpw}. 
Concerning its l.h.s., $\widehat{\alpha}_T$ can be obtained from a lattice simulation at any $\beta$ and $\mu_l$, after 
properly shifting the lattice momentum $a q$ through multiplication by the ratio of lattice spacings 
$a(\beta_0,\mu^0_l)/a(\beta,\mu_l)$, being so left with the physical momentum expressed in units of $a^{-1}(\beta_0,\mu^0_l)$, 
$k_L$. Therefore, the ratio $a(\beta_0,\mu^0_l)/a(\beta,\mu_l)$ and $c_{a2p2}$ can be obtained by requiring the data 
from l.h.s. to be best fitted by the r.h.s., for any $k_L > 0.5$, with the minimum-$\chi^2$  criterion.  
The coefficient $c_{a2p2}$ is treated as a free parameter although, by dimensional arguments from Eqs.~(\ref{eq:a2p2},\ref{eq:1overpx}), it could be thought not to depend on the lattice spacing and, hence, 
not to depend on $\beta$ and $\mu_l$ either. However, as $c_{a2p2}$ is to be 
determined by a fit of \eq{eq:1overpx} to lattice data, it might 
borrow something from higher-order lattice artefacts and its fitted values might slightly depend on the lattice parameters 
for the simulation. The procedure has been followed for the analysis of data from the three ensembles at $\beta=1.90$ and 
the fit results have been shown in Tab.~\ref{tab:matching}. As can be seen there, we are left with high-quality fits 
where, as expected, the coefficients $c_{a2p2}$ are fully compatible with each other and, 
together with those obtained for $\beta=2.10$ in the previous section, all happen to lie fairly in the same ballpark. 
In previous analysis~\cite{Blossier:2011tf,Blossier:2012ef}, we {\it imposed} $c_{a2p2}$ to be the same for different simulations at any $\beta$ and $\mu_l$, mainly in order to stabilize the fits; here we {\it found} this to happen within a reasonable degree of approximation which is compatible with assuming the extraction for $c_{a2p2}$ to be polluted by higher-order artefacts. 
This strongly supports the approach followed in refs.~\cite{Blossier:2011tf,Blossier:2012ef} and confirms the reliability of previous and present results.

\begin{table}[ht]
\begin{tabular}{|c|c|c|c|}
\hline 
$\mu_l$ & $a(2.10,0.002)/a(1.90,\mu_l)$ & $c_{a2p2}$ & $\chi^2/$d.o.f. 
\\
\hline
0.003 &  0.6798(74) & -0.0076(6) & 20.2/82 
\\
\hline 
0.004 & 0.6683(72) & -0.0067(6) & 14.4/82
\\
\hline 
0.005 & 0.6775(73) & -0.0081(6) & 45.7/82
\\ 
\hline
\end{tabular}
\caption{\small Best-fit parameters obtained with \eq{eq:scaling2} and the data for the three ensembles at $\beta=1.90$. 
The fitting momenta window is defined by $k_L>0.5$, as explained in the text.
}
\label{tab:matching}
\end{table}

\subsection{Chiral extrapolation and lattice calibration}

\begin{figure}[ht]
  \begin{center}
    \includegraphics[width=12cm]{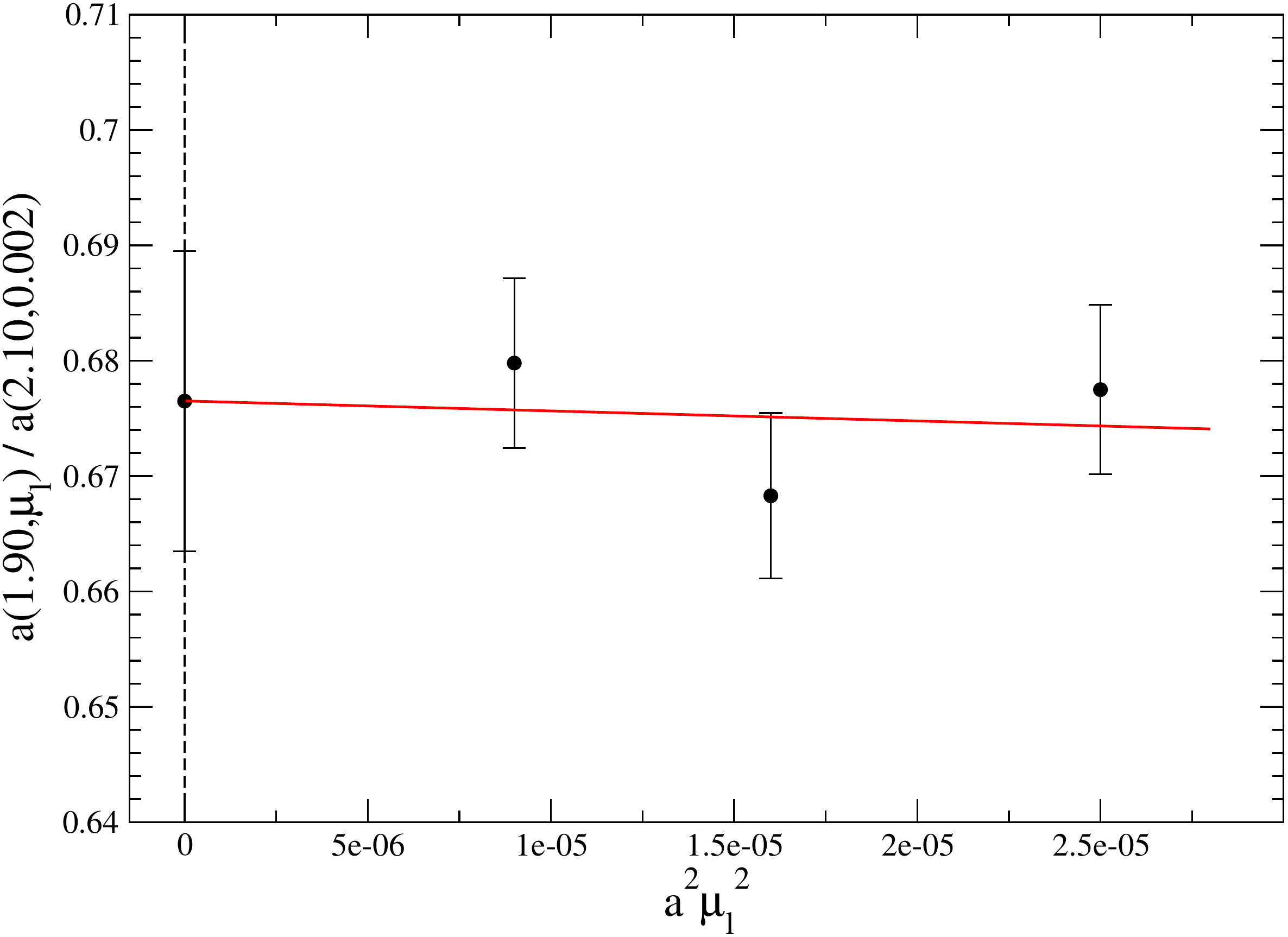}
  \end{center}
\caption{\small ratios of lattice spacings from Tab.~\ref{tab:matching} plotted in terms of $a^2\mu^2_l$ which, in the case of fermions simulated with a twisted-mass action at the maximal twist, is proportional to the light quark mass squared. The solid red line corresponds to 
an extrapolation to the chiral limit according to \eq{eq:aml}.
}
\label{fig:aml}
\end{figure}

In the previous subsection, we have obtained the ratio of the lattice spacing for the simulations at $\beta=1.90$ and three 
different dynamical flavour masses, $a(1.90,\mu_l)$, over the spacing for the simulation at $\beta=2.10$ and $\mu_l=0.002$ (see 
Tab.~\ref{tab:matching}). Since the twisted-mass fermion action has been tuned to the maximal twist angle, 
the flavour mass happens to be proportional to $\mu_l$. On the other hand, all the flavour mass dependence is assumed to be captured by the lattice spacing which should be expected to behave 
linearly on $a^2\mu^2_l$ since the lattice action is improved at the ${\cal O}(a)$-order. One can thus write:
\beq\label{eq:aml}
\frac{a(2.10,0.002)}{a(\beta,\mu_l)} \ = \ \frac{a(2.10,0.002)}{a(\beta,0)} \ 
\left( \rule[0cm]{0cm}{0.6cm} 1 + \ c_\mu \ a^2(\beta,0) \mu^2_l \ + \ o(a^2\mu^2_l) \right) \ ,
\eeq
according to which, the ratios from Tab.~\ref{tab:matching} can be extrapolated to reach the chiral limit, when
the light flavour mass is exactly put to zero. Thus, as can be seen in Fig.~\ref{fig:aml}, 
we obtain a remarkably weak slope: 
\beq
c_\mu \ = \ -1.1(7.8) \times 10^2 \ ,
\eeq
clearly compatible wity zero, and 
\beq\label{eq:Phy}
a(2.10,0.002) \ = \ 0.677(13) \ a(1.90,0) \ = \ 0.0599(27) \ \mbox{\rm fm} \ ;
\eeq
where we applied the very recent result $a(1.90,0)=0.0885(36)\ {\rm fm}$~\cite{Silvano:2013nw}, obtained by the ETM collaboration 
through chiral fits for the lattice pseudoscalar masses and decay constants that are required 
to take the experimental $f_\pi$ and $m_\pi$ at the extrapolated physical point. 
The pion masses for the simulations happen to range from 270 to 510 MeV.  

\begin{figure}[ht]
  \begin{center}
  	\begin{tabular}{cc}
    \includegraphics[width=10cm]{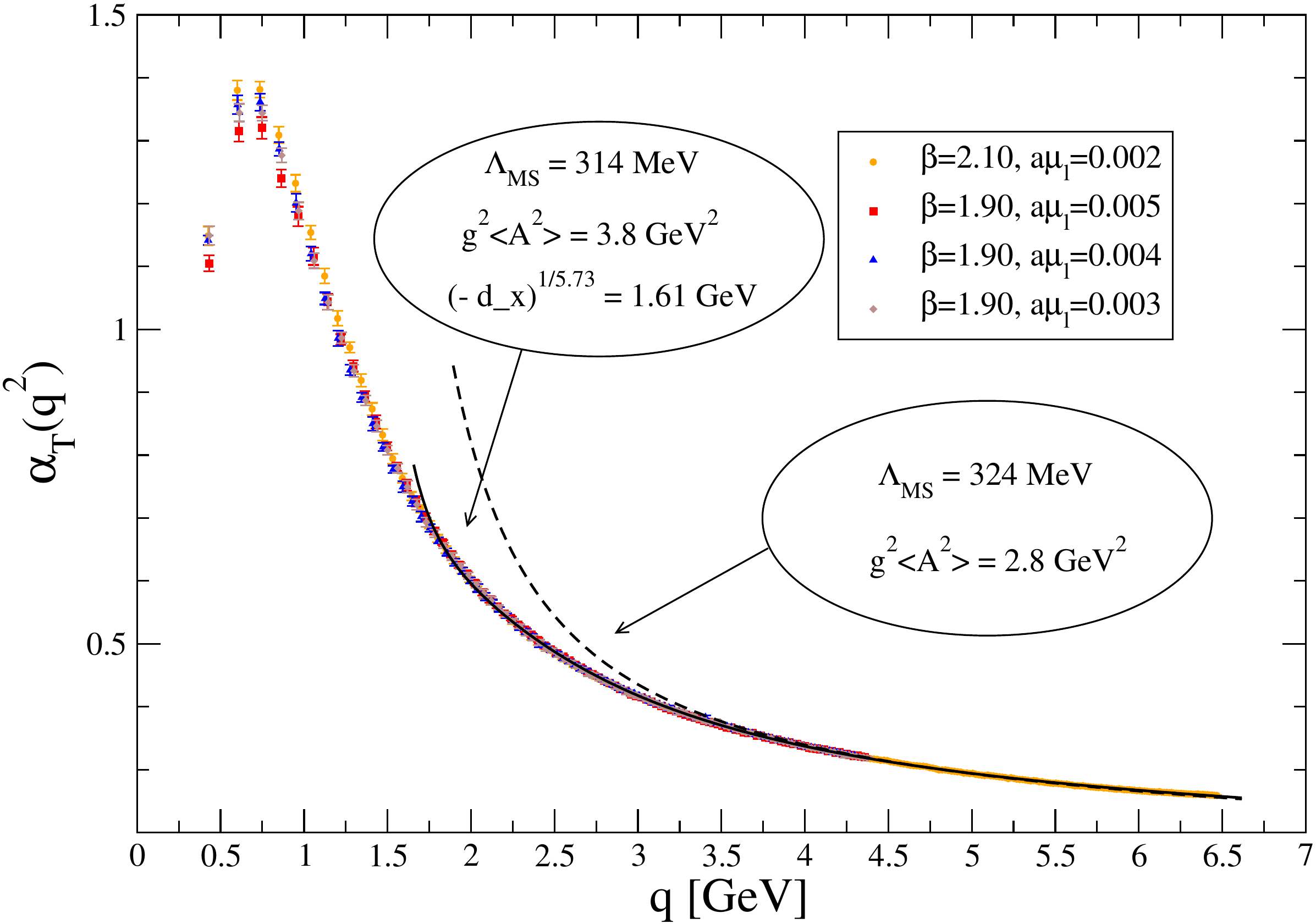} & 
    \includegraphics[width=7.5cm,height=7cm]{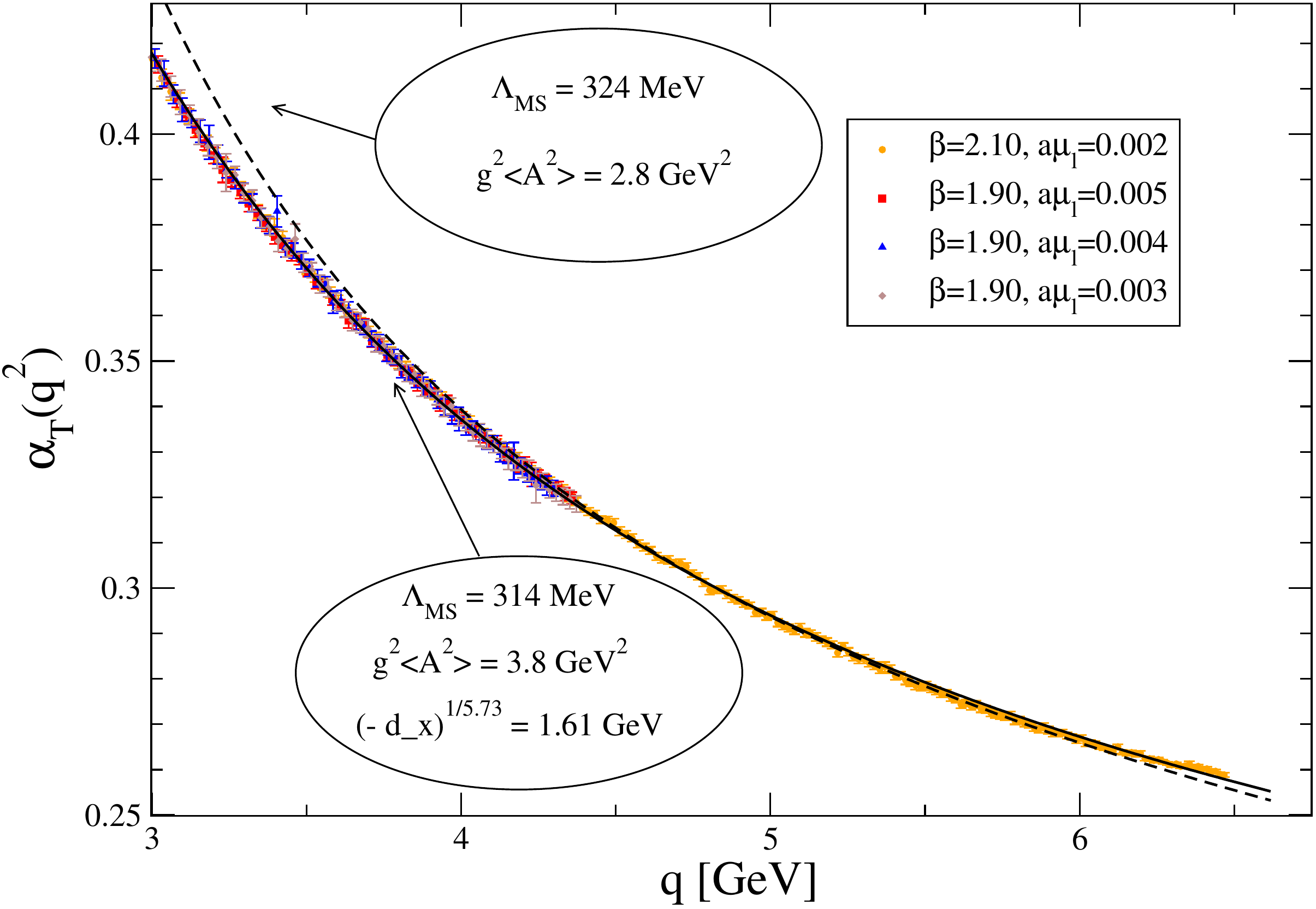}
	\end{tabular}	
  \end{center}
\caption{\small The Taylor running coupling obtained from the lattice after the appropriate removal of 
$O(4)$-breaking ($H(4)$-extrapolation) and $O(4)$-invariant artefacts. The curves 
are fitted to the physical running defined by \eq{eq:atA2} with the parameters shown in Tab.~\ref{tab:Phy}, 
the solid (dashed) black line corresponds to the large (small) fitting window. The right plot is  
a zoom, focussing on the large momenta region, of the left one.
}
\label{fig:Phy}
\end{figure}

Now, \eq{eq:Phy} can be used to make a conversion of our results to physical units and, together with the ratios of Tab.~\ref{tab:matching}, the lattice estimates of $\widehat{\alpha}_T$ from simulations at $\beta=2.10$ and $\beta=1.90$ 
of Fig.~\ref{fig:raw} can be cured for $O(4)$-invariant lattice artefacts from \eq{eq:1overpx} and plotted in terms of the 
momentum in physical units, where the evidence for the scaling expressed by \eq{eq:scaling} is to be weighted through  
the superposition of data from different simulations. This can be seen in Fig.~\ref{fig:Phy}, while the fit results in physical 
units are reported in Tab.~\ref{tab:Phy}.

\begin{table}[ht]
\begin{tabular}{||c|c|c|c||}
\hline
fit window [GeV] & $\Lambda_{\msbar}$ [MeV] & $g^2 \langle A^2 \rangle$ [GeV$^2$]& $(-d_x)^{1/x}$ [GeV] \\
\hline
\hline
[4.3,6.0] & 324(18) & 2.8(8) & \\
\hline
[1.7,6.6] & 314(16) & 3.8(6)  & 1.61(7)\\
\hline
\end{tabular}
\caption{Fitted parameters from previous sections, now expressed in physical units with the help of \eq{eq:Phy}. The error estimates include the uncertainty on the lattice spacing determination given in \eq{eq:Phy}.
}
\label{tab:Phy}
\end{table}

\section{$\boldsymbol{\amsbar{q^2}}$ from the ghost-gluon coupling}

The $\msbar$ running coupling can be obtained by the integration 
of the $\beta$-function, with the coefficients now in $\msbar$-scheme, and $\Lambda_{\overline{\rm MS}}$ 
both for ${\rm N}_f=4$. We will read $\Lambda_{\overline{\rm MS}}$ from Tab.~\ref{tab:Phy},
\beq\label{eq:LMS}
\Lambda_{\msbar} \ = \ 314(7)(14){\bf(10)} \times \left( \frac{0.0599 \ \mbox{\rm fm}}{a(2.10,0.002)} \right) 
 \ \mbox{\rm MeV}
\eeq
where we take the central value from the fit including a higher-order power correction and estimate a 
conservative systematic error (in bold faced characters) with its difference from the value obtained with the fit over 
the small window, without including the higher-order power. The two other errors quoted in \eq{eq:LMS} correspond 
to the statistical uncertainty from the fit (first) and from the lattice spacing given by \eq{eq:Phy} (second).

Then, we can apply the result in \eq{eq:LMS} to run the coupling down to the scale of the $\tau$ mass, 
below the bottom quark mass threshold,  and compare the result with the estimate from 
$\tau$ decays~\cite{Beringer:1900zz,Bethke:2011tr}, $\alpha_{\msbar}(m_\tau^2)=0.330(14)$. 
This gives, with the 1-$\sigma$ propagation given by 
\beq
\sigma^2\left(\amsbar{q^2}\right) \ = \ 4 \beta^2_{\msbar}\left(\amsbar{q^2}\right) \ \frac{\sigma^2\left(\Lambda_{\msbar}\right)}{\Lambda^2_{\msbar}} \ 
\eeq
for each error contribution in \eq{eq:LMS}, the following result at the $\tau$-mass scale: 
\beq\label{eq:amtau}
\amsbar{m_\tau^2} \ = \ 0.336(4)(8)(6) \ ,
\eeq
in good agreement with the one from $\tau$ decays.

In order to determine $\amsbar{q^2}$ at the $Z^0$ mass scale, we should first 
run the coupling up to the $\msbar$ running mass for the bottom quark, $m_b$, 
with $\beta$-coefficients and $\Lambda_{\msbar}$ estimated for 4 quark flavours, 
apply next the matching formula~\cite{Nakamura:2010zzi}:
\beq
\ams^{N_f=5}(m_b) = \ams^{N_f=4}(m_b) 
\left( 1 + \sum_n c_{n0} \left(\ \ams^{N_f=4}(m_b)\right)^n \right) ,
\eeq
where the coefficients $c_{n0}$ can be found in ref.~\cite{Chetyrkin:2005ia,Schroder:2005hy} and 
finally run from the bottom mass up to the $Z^0$ mass scale with $\beta$-coefficients for 
$N_f=5$. We obtain:
\beq\label{eq:amz0}
\ams(m_Z^2)=0.1196(4)(8)(6) \ ,
\eeq
where we have again propagated all the error contributions from \eq{eq:LMS}. This result is compatible with the last lattice results averaged by PDG~\cite{Beringer:1900zz}, $0.1185(7)$, and with its world average without including lattice results, 
$0.1183(12)$. Eqs.~(\ref{eq:amtau},\ref{eq:amz0}) 
update the results of \cite{Blossier:2012ef} with a much higher statistics for our sample of gauge field 
configurations which allowed for a more precise result and a more reliable error analysis. The latter is mainly because, 
in order to be left with stable fits, we did not need to combine data from simulations at different $\beta$ and made 
no hypothesis on how the lattice artefacts scale for different lattice spacings.

\section{Conclusions}

The strong coupling renormalized in the MOM  Taylor scheme has been obtained from ghost and gluon 
propagators that have been in their turn computed from several high-statistical samples of gauge field 
configurations, simulated at two different bare couplings, $\beta$'s, and four different 
sets of masses for the degenerate up and down and non-degenerate strange and 
charm twisted-mass dynamical quark flavors. We thus updated our previous results by performing 
a  significatively improved analysis, which reduced the statistical errors and allowed for a better control 
of the systematic uncertainties, as those related with higher-order OPE corrections and lattice artefacts. 
Furthermore, we took full account within the estimate of statistical errors of the matching procedure to 
determine the ratio of different lattice spacings and of the chiral extrapolation to the physical point 
for their absolute ``calibration". Last but not least, we avoided to assume the constancy of the coefficient 
$c_{a2p2}$ for the $O(4)$-invariant lattice artefacts and indeed found it to happen consistently with 
the approximation of neglecting higher-order artefacts. 

As a result of our improved analysis, we confirmed the need to include nonperturbative corrections 
in order to describe the lattice data for the MOM Taylor strong coupling for momenta below 7 GeV. 
After dealing properly with lattice artefacts, the Wilson coefficient for the leading OPE power correction, 
known at the ${\cal O}(\alpha^4)$-order, is found to account for data with momenta roughly above 4 GeV 
and, along with an effective higher-order power correction, above $\simeq 2$ GeV.  The value of 
$\Lambda_{\msbar}$, for $N_f$=4, has been obtained from the running of the 
Taylor strong coupling and used then to estimate the $\msbar$ coupling at the $\tau$-mass 
scale and, properly handling the transition from $N_f$=4 to $N_f$=5, at the $Z^0$-mass scale. 
In both cases, our estimates agree with the ``{\it world average}" results the Particle Data Group provides. 
Therefore, with the lattice-regularized QCD action we used, the pion mass and decay constant have been taken 
as the physical scales to size the strong coupling in the appropriate momentum window, roughly from 2 to 7 GeV.
As far as perturbative QCD is, in its turn, applied to run the coupling from those momenta up to $Z^0$-mass scale, a main conclusion of this work is that {\it QCD is successfully bridging from 
the low-momentum pion sector up to the very UV domain for the strong interactions}.

\section*{Acknowledgements} 
We are particularly indebted to A. Le Yaouanc, J. P. Leroy 
and J. Micheli for participating in many fruitful discussions at 
the preliminar stages of this work. 
We thank the support of Spanish MICINN FPA2011-23781 research project and  
the IN2P3 (CNRS-Lyon), IDRIS (CNRS-Orsay), TGCC (Bruyes-Le-Chatel), CINES (Montpellier). 
K.Petrov is part of P2IO Laboratory of Excellence.

\appendix 

\section{Hypercubic $H(4)$-extrapolation}
\label{sec:apH4}  
     
The first kind of artefacts that can be systematically 
cured~\cite{Becirevic:1999uc,deSoto:2007ht} are those due to the
breaking of the  rotational symmetry of the Euclidean space-time when using an
hypercubic lattice,  where this symmetry is restricted to the discrete $H(4)$
isometry group. Let us consider a general dimensionless lattice quantity 
$Q(a q_\mu)$ that will be conveniently averaged over every orbit of 
the group $H(4)$.  In general several orbits of $H(4)$ correspond to one value of $q^2$.
Defining the $H(4)$ invariants
 \beq
 q^{[4]}=\sum_{\mu=1}^{4} q_\mu^4\qquad q^{[6]}=\sum_{\mu=1}^{6} q_\mu^6 \ ,
 \eeq
it happens that the orbits of $H(4)$ are 
labelled\footnote{On totally general grounds, any $H(4)$-invariant 
polynome can be written only in terms of the four 
invariants $q^{[2 i]}$ with $i=1,2,3,4$~\cite{Becirevic:1999uc,deSoto:2007ht}. 
As a consequence of the upper cut for momenta, the first three of these invariants suffice 
to label all the orbits we deal with and hence any presumed dependence on $q^{[8]}$ is  
neglected.} 
by the set $q^2, a^2 q^{[4]},
a^4 q^{[6]}$.  We can thus define the quantity $Q(a q_\mu)$ averaged over 
$H(4)$ as
\beq\label{Q246}
Q(a^2\,q^2, a^4q^{[4]}, a^6 q^{[6]}).
\eeq
In the continuum limit the effect of $ a^2 q^{[4]},
a^4 q^{[6]}$ vanishes, as indeed happens for 
a free lattice propagator:
\beq
\frac 1 {\displaystyle \frac1 {a^2} \sum_\mu \left( 1- \cos(a q_\mu) \right)^2} 
 \ = \ 
 \frac 1 {q^2} \ \left(1 + \frac 1 {12} \frac{a^2 q^{[4]}}{q^2} + \cdots \right) \ .
\eeq
If the lattice spacing is small enough such that $\epsilon=a^2 q^{[4]}/q^2 \ll 1$,
the dimensionless lattice correlation function defined in~\eq{Q246} can be
expanded in powers of $\epsilon$:
\beq\label{eq:apQ1}
Q(a^2\,{q}^2, a^4q^{[4]}, a^6 q^{[6]})
= \widehat{Q}(a^2q^2) +
\left.\frac{dQ}{d\epsilon}\right|_{\epsilon=0} a^2
\frac{q^{[4]}}{q^2} + \cdots
\eeq
Then,  one can  fit the
coefficient $dQ/d\epsilon$ 
from the whole set of orbits sharing the same $q^2$ and get $ \widehat{Q}$, the extrapolated value of $Q$
free from $H(4)$ artefacts. If we further assume that the coefficient 
\[
\left.R(a^2q^2) = 
\frac{dQ}{d\epsilon}\right|_{\epsilon=0}
\]
has a smooth dependence on $a^2q^2$ over a given momentum window, we can 
expand it as $R=R_0+R_1 a^2q^2$ and make a global fit in the whole 
momentum window to extract the extrapolated value
of $Q$ for any momenta $q$. 
The contribution from the term $a^4 q^{[6]}/q^2$ has been also neglected. The very good matching 
for the estimates of $\alpha_T$ from simulations with different lattice spacings we have obtained (and that 
can be seen in Fig.~\ref{fig:Phy}) proves the latter to be a good approximation.

On the other hand, the continuum limit (and the full $O(4)$ rotational invariance) has to be 
reached when $\epsilon=a^2 q^{[4]}/q^2 \to a^2 q^2/2 \to 0$. Then, we could have also 
expanded the attice correlation function defined in~\eq{Q246} as
\beq\label{eq:apQ2}
Q(a^2\,{q}^2, a^4q^{[4]}, a^6 q^{[6]})
= \widehat{Q'}(a^2q^2) +
\left.\frac{dQ}{d\epsilon}\right|_{\epsilon=0} 
\left( a^2 \frac{q^{[4]}}{q^2} - \frac{a^2 q^2}{2} \right) + \cdots \ .
\eeq
Anyhow, $\widehat{Q}$ in \eq{eq:apQ1} only differs from $\widehat{Q'}$ in \eq{eq:apQ2} by 
a term proportional to $a^2q^2$, not breaking $O(4)$ symmetry, that will be furtherly dropped 
in the next step (see \eq{eq:a2p2}), when curing the remaining $O(4)$-invariant artefacts.
Then, we have chosen to apply \eq{eq:apQ1} that, when $\alpha_T$ replaces $Q$, reads as \eq{eq:H4}.

It is finally worthwile to mention that we considered in this work  anisotropic lattice
of the type $L^3$x$T$, with $T=2L$.  This finite volume effect reduces the $H(4)$
lattice symmetry to $H(3)$.  Deviations from $H(4)$  are to be expected in the
long-distance physics. But ultraviolet physics should not be affected.  As far
as we are interested in the high-momentum regime,  we will assume the previous
treatement of the lattice artefacts to be valid.



\end{document}